%% file: main.tex
\documentclass[sigplan,nonacm]{acmart}

\AtBeginDocument{%
  }

\setcopyright{none}
\renewcommand\footnotetextcopyrightpermission[1]{}

\usepackage[utf8]{inputenc}
\usepackage{amsmath,amsthm}
\usepackage{MnSymbol}
\usepackage{cleveref}
\usepackage{listings, fancyvrb, multicol}
\usepackage{enumitem}
\usepackage{float}
\usepackage{tikz-cd}
\usepackage{microtype}
\usepackage{multirow}
\usepackage{listings}
\usepackage{comment}
\usepackage{xspace}
\usepackage{graphicx, graphics}
\usepackage{subcaption}
\usepackage{svg}
\usepackage{balance}
\usepackage{etoolbox}

\usepackage{bm}

\usepackage[table]{xcolor}

\newcommand{\mosaic}{{Mosaic}\xspace}
\newcommand{\mosaicsched}{{MosaicSched}\xspace}

\newcommand*\circled[1]{\tikz[baseline=(char.base)]{
            \node[shape=circle,draw,inner sep=2pt] (char) {#1};}}

\begin{document}

\title[\mosaic]{\mosaic: GPU Sharing with Latency Guarantees through Kernel-Level Interference Prediction}

\author{Foteini Strati}
\affiliation{%
  \institution{ETH Zurich}
  \country{Switzerland}}

\author{Ethan Graham}
\affiliation{%
  \institution{ETH Zurich}
  \country{Switzerland}}

\author{Leo Stephan}
\authornote{Work done while at ETH Zurich}
\affiliation{%
  \institution{University of Oxford}
  \country{United Kingdom}}

\author{Paul Elvinger}
\authornotemark[1] 
\affiliation{%
  \institution{Wayve}
  \country{United Kingdom}}

\author{Ana Klimovic}
\affiliation{%
  \institution{ETH Zurich}
  \country{Switzerland}}

\begin{abstract}

GPUs are increasingly in demand for AI workloads, yet often remain substantially underutilized, motivating workload colocation. However, colocation introduces interference that can degrade latency-critical workloads. Existing approaches mitigate interference using either heuristic-based scheduling or interference predictors. Heuristics rely on coarse-grained metrics that overlook important interference sources, while predictors often depend on simulator-specific or similarly coarse metrics. However, GPU interference is complex and arises from multiple mechanisms, including thread-block placement, memory hierarchy contention, and intra-SM resource contention. We present \textit{\mosaic}, a kernel-level interference predictor that explicitly models these mechanisms using a combination of analytical and lightweight learned models. Across four GPU architectures, \mosaic reduces prediction error by up to an order of magnitude compared to prior predictors. 
We integrate \mosaic into a scheduler, \textit{\mosaicsched}, that performs online kernel admission control and selects between full-GPU colocation and SM partitioning to maximize best-effort throughput while satisfying latency SLOs. Across all workloads, \mosaicsched keeps the p99 latency
below or very close to the target SLO.

\end{abstract}

\maketitle

\input{content/intro}
\input{content/background}

\input{content/motivation}

\input{content/kernelpredictor}

\input{content/workloadpredictor}

\input{content/eval}
\input{content/discussion}
\input{content/conclusion}

\newpage
\bibliographystyle{ACM-Reference-Format}
\bibliography{references}

\input{content/appendix}

\end{document}

%% file: content/intro.tex
\section{Introduction}

Graphics Processing Units (GPUs) are widely used to train and serve modern AI models. As AI adoption continues to grow, so does the demand for GPU resources. Paradoxically, despite their high demand, GPUs are often substantially underutilized.  A week-long study of a production ads inference service at Meta shows that the GPU utilization remained between 20\%-40\% over the entire period~\cite{Coppock25Lithos}. Microsoft reports less than or equal to 50\% GPU utilization across 400 training and inference jobs~\cite{Gao24Microsoft}. GPU underutilization can arise from various factors, such as communication and data-processing overheads, checkpointing, stragglers, small batch sizes, and varying resource demands across a workload~\cite{Gujarati20Clockwork, Mingcong22Reef, strati25pccheck,  Strati24Orion, Zhu25Nanoflow}. Such underutilization results in significant cost and energy inefficiencies, as GPUs that are not fully utilized, still consume a significant amount of power, often quite close to their Thermal Design Power (TDP)~\cite{lei2026energyidle}.

\begin{figure}[htbp]
    \centering
    \includegraphics[width=\columnwidth]{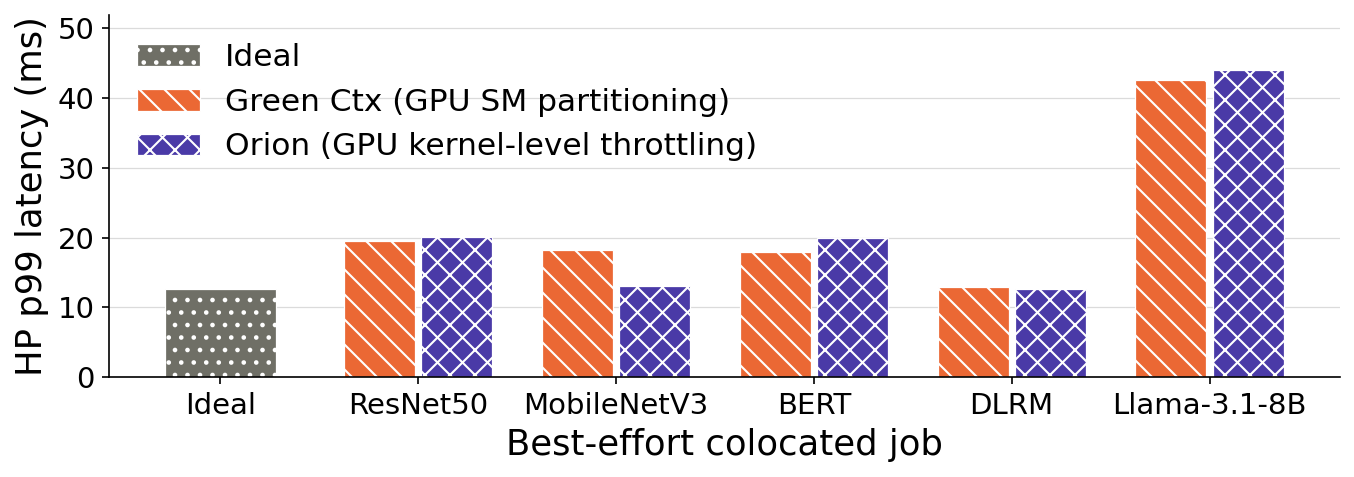}
    \caption{P99 latency of ResNet50 inference as a high-priority job, colocated with different best-effort jobs under Orion and Green-Context-based scheduling. Current GPU scheduling approaches try to minimize interference, but cannot provide performance guarantees for latency-critical jobs.
    }
    \label{fig:orion-rnet8}
\end{figure}

A common approach to improve GPU utilization is to spatially colocate multiple workloads on the same GPU, allowing them to execute concurrently. However, concurrent execution introduces contention for shared GPU resources, which can significantly degrade application performance. 
This has motivated the design of GPU schedulers that control the colocation of latency-critical and best-effort workloads~\cite{Strati24Orion, Coppock25Lithos, Mingcong22Reef, hu2026hummingbird}.

There are two main approaches that GPU schedulers use to try to minimize interference. The first is relying on heuristics to guide scheduling decisions, using coarse resource characteristics such as Streaming Multiprocessor (SM) occupancy, memory bandwidth, or overall compute and memory resource usage~\cite{Mingcong22Reef, Sudipta24Usher, Strati24Orion}. While effective in some cases, these heuristics capture only a subset of GPU interference mechanisms and can therefore lead to suboptimal scheduling decisions~\cite{Elvinger26Interference}. For example, Figure~\ref{fig:orion-rnet8} shows the p99 latency of a high-priority job (ResNet50 inference) running under different colocation scenarios with best-effort jobs using the Orion scheduler~\cite{ Strati24Orion} and a Green-Context-based scheduler that partitions the GPU SMs across the jobs and allocates more GPU SMs to the high-priority (HP) job. Neither approach preserves the high-priority job's performance across all colocation scenarios: p99 latency can be as high as 3.4× compared to the ideal (no-colocation) scenario. 

The second approach is explicitly predicting the slowdown caused by colocation and using these predictions to guide scheduling~\cite{Zhao20HSM, Zhao19Themis, Tan23Gpupool, Han24Kace, Zhang25Bless}. While this should enable more informed scheduling decisions, it is challenging to accurately predict interference. Today's predictors are inaccurate because they either rely on simulator-specific metrics that do not generalize in real GPUs~\cite{Zhao20HSM, Zhao19Themis, Tan23Gpupool} or they use coarse-grained hardware statistics that fail to capture the diverse sources of GPU interference~\cite{Han24Kace}.

The above mechanisms are suboptimal because GPU interference is a complex phenomenon. GPU interference arises from multiple distinct mechanisms—including thread-block placement, memory hierarchy contention, and
intra-SM resource contention—that existing heuristics and
predictors either ignore or conflate~\cite{Elvinger26Interference}. Accurately modeling these mechanisms is essential for practical interference-aware GPU scheduling. 

We present \mosaic, a kernel-level
interference predictor that explicitly decomposes GPU interference into its different sources and models each separately. \mosaic enables GPU schedulers to provide guarantees such as p99 latency SLOs. The \mosaic predictor design is guided by three key observations. First, accurately predicting slowdown requires capturing how kernels overlap in space and time, which in turn requires modeling thread-block placement across SMs. Second, aggregate hardware metrics such as memory bandwidth are often insufficient, as slowdown depends on finer-grained properties such as memory access behavior and latency-hiding capability. Third, while some interference mechanisms can be modeled analytically, others are better captured with lightweight learned models. \mosaic combines both to predict slowdown for kernels running on default CUDA streams or Green-Context-based streams that partition GPU SMs. Across four GPU types (H100, A100, V100, and RTX-3090), Mosaic reduces prediction error by up to an order of magnitude compared to prior predictors.

Using the \mosaic predictor, we build \mosaicsched, a GPU scheduler that performs online admission control for best-effort kernels, in order to maintain p99 latency guarantees for the high-priority job. For every kernel submitted by the best-effort job, \mosaicsched queries the predictor to get the estimated slowdown on the colocated high-priority kernel and admits the best-effort kernel only if and when the predicted end-to-end latency of the high-priority workload remains within a user-specified service level objective (SLO).
\mosaicsched also optimizes performance by automatically selecting between two colocation modes: 1) full-GPU colocation, where both workloads can use all SMs and therefore dynamically share GPU resources, and 2) SM partitioning with CUDA Green Contexts~\cite{cudagreencontexts}, where each workload is restricted to a dedicated subset of SMs, reducing interference at the cost of limiting available resources. Across all workloads and SLO targets, \mosaicsched keeps the p99 latency below or very close to the target SLO.

Our experiments also reveal important insights about GPU sharing.
First, neither colocation mode is universally optimal. The tradeoff between full-GPU and SM partitioning colocation depends on the resource characteristics of both the high-priority and best-effort workloads. While full-GPU execution offers more resources to best-effort kernels, it can also lead to stronger interference, reducing admission opportunities for the best-effort kernels. Conversely, SM partitioning reduces interference but limits the throughput attainable by the admitted best-effort kernels, and the optimal partition depends strongly on the workloads' compute and memory characteristics. 
Second, maximizing GPU utilization does not necessarily minimize deployment cost. While beneficial for many workload pairs, GPU sharing may require throttling the best-effort workload so extensively to meet the high-priority latency target that its completion time increases substantially, eliminating the cost advantage of sharing a single GPU.


%% file: content/background.tex
\section{Background}\label{background}

We describe the architecture of modern GPUs, focusing on NVIDIA GPUs and using NVIDIA/CUDA terminology.

\textbf{GPU architecture:}
NVIDIA GPUs consist of Streaming Multiprocessors (SMs), grouped into Texture Processing Clusters (TPCs). Each SM contains several subpartitions (SMSPs), each with a warp scheduler, register file, and execution pipelines (e.g., FMA, Tensor Cores, LSU). Each SM has configurable on-chip memory that can be divided between L1 cache and shared memory (the \textit{L1/shared-memory carveout}), while all SMs share the L2 cache and off-chip memory.

\textbf{GPU programming model:} Users submit work to the GPU by launching kernels, and specifying the number of blocks (grid size) and the number of threads (block size) for each kernel. Kernels and memory operations are submitted through streams, which enforce ordering among operations within a stream while allowing operations from different streams to execute concurrently. By default, a kernel's blocks can execute on any available SM. CUDA Green Contexts allow a stream to be associated with a subset of SMs, restricting its kernels to that partition~\cite{cudagreencontexts}.

\textbf{Thread scheduling:} Thread blocks are assigned to SMs and remain on the same SM until completion. The number of resident blocks per SM is limited by available threads, registers, and shared memory. When a kernel has more blocks than can simultaneously reside on the GPU, its execution proceeds in successive \textit{waves}, where each wave consists of the blocks that can execute concurrently given these resource constraints. 
A GPU executes kernel threads at the granularity of warps, typically consisting of 32 threads. Once a block is scheduled on an SM, its warps are assigned to one
of the subpartitions. At each clock cycle, the
warp scheduler in each subpartition chooses one eligible warp and schedules one or more instructions from that warp. A warp is considered \textit{eligible} if its instruction has been fetched, it has no outstanding dependencies, and the execution unit required by the instruction is available. If a warp is not in that state, it is considred \textit{stalled}. A warp can stall for multiple reasons, e.g long and short scoreboard stalls (global and shared memory accesses), branch resolution, synchronization barriers, and execution pipeline contention~\cite{ncuwarpstallreasons}.

\textbf{Memory subsystem and access:} 
Shared memory is private to each SM, allocated per thread block, and organized into banks. Bank conflicts serialize accesses. Global memory is accessed through per-SM L1 caches and a shared L2 cache. Memory requests are served in 32-byte sectors. Accesses from threads within a warp are coalesced into as few transactions as possible~\cite{coalescingcuda}.

%% file: content/motivation.tex
\section{Why existing GPU schedulers are not enough}

\begin{table}[tp!]
\footnotesize
\centering
\setlength{\tabcolsep}{3pt}
\renewcommand{\arraystretch}{1.15}

\begin{tabular}{|p{0.16\columnwidth}|
                >{\centering\arraybackslash}p{0.135\columnwidth}|
                >{\centering\arraybackslash}p{0.08\columnwidth}|
                >{\centering\arraybackslash}p{0.115\columnwidth}|
                >{\centering\arraybackslash}p{0.105\columnwidth}|
                >{\centering\arraybackslash}p{0.085\columnwidth}|
                >{\centering\arraybackslash}p{0.13\columnwidth}|}
\hline
\centering\textbf{System}
& \multicolumn{3}{c|}{\textbf{Inter-SM}}
& \multicolumn{3}{c|}{\textbf{Intra-SM}}
\\ \hline

& Thread Block Scheduler
& L2 Cache
& Memory Bandwidth
& Warp Scheduler
& CUDA cores
& L1 cache / Shared Memory
\\ \hline

Laius~\cite{Wei19Laius}
& X & X & $\checkmark$ & X & X & X \\ \hline

GPUlet~\cite{Choi22Gpulet}
& X & $\checkmark$ & $\checkmark$ & X & X & X \\ \hline

iGniter~\cite{xu2022igniter}
& X & $\checkmark$ & X & $\checkmark$ & X & X \\ \hline

Reef~\cite{Mingcong22Reef}
& $\checkmark$ & X & X & X & X & X \\ \hline

Orion~\cite{Strati24Orion}
& $\checkmark$ & X & $\checkmark$ & X & $\checkmark$ & X \\ \hline

Usher~\cite{Sudipta24Usher}
& $\checkmark$ & X & $\checkmark$ & X & X & X \\ \hline

LithOS~\cite{Coppock25Lithos}
& $\checkmark$ & X & X & N/A & N/A & N/A \\ \hline

\textbf{Mosaic} & $\bm{\checkmark}$ & $\bm{\checkmark}$ & $\bm{\checkmark}$ & $\bm{\checkmark}$ & $\bm{\checkmark}$ & $\bm{\checkmark}$ \\ \hline

\end{tabular}

\caption{Schedulers for GPU colocation across workloads. As seen in the
table, no scheduler can accurately capture all main sources of GPU
interference. Works such as LithOS~\cite{Coppock25Lithos} partition TPCs
so intra-SM interference is not relevant.}
\label{table:schedulers}
\end{table}

Despite the growing demand for high-end GPUs, production clusters consistently exhibit low GPU utilization, leading to substantial cost and energy inefficiencies~\cite{Coppock25Lithos, Gao24Microsoft, Xiao20Antman, lei2026energyidle}. 
Spatial colocation of GPU kernels is a promising way to improve utilization, both across workloads and within a workload. Multiple works explore inter-workload GPU colocation (especially for inference jobs), since kernels from separate jobs are independent, and can potentially execute concurrently~\cite{Coppock25Lithos, Sudipta24Usher, Strati24Orion, Mingcong22Reef, xu2022igniter}. Distributed training and inference systems commonly use intra-workload colocation to overlap communication and computation~\cite{Chen24Centauri, zhang2025comet, deepep2025}, and recent LLM inference frameworks such as NanoFlow~\cite{Zhu25Nanoflow} and Nexus~\cite{shi2025nexus} concurrently schedule compute-, memory-, and network-bound operators as well as prefill and decode phases. However, concurrent execution can cause interference as kernels contend for shared GPU resources, including SMs and the memory hierarchy~\cite{Elvinger26Interference}, potentially negating the benefits of colocation by increasing execution times for both intra- and inter-job settings~\cite{shi2025nexus, Coppock25Lithos, Strati24Orion},  as shown in Figure ~\ref{fig:orion-rnet8}. Consequently, effective GPU scheduling depends on accurately modeling kernel interference, which remains challenging for existing approaches.

\subsection{Heuristic-based GPU schedulers}


Table~\ref{table:schedulers} summarizes existing GPU schedulers and indicates which sources of inter-SM and intra-SM interference they consider. No scheduler addresses all sources of GPU interference. For example, Usher~\cite{Sudipta24Usher} colocates workloads based on their achieved occupancy and memory bandwidth usage, while iGniter\cite{xu2022igniter}~ finds pairs for colocation based on their L2 cache usage. Reef~\cite{Mingcong22Reef} schedules kernels based on their SM demand, while Orion~\cite{Strati24Orion} additionally schedules kernels with complementary resource requirements.
However,  as Elvinger et al.~\cite{Elvinger26Interference} show, GPU interference arises from multiple sources, making single-metric heuristics insufficient. Even LithOS~\cite{Coppock25Lithos}, which partitions TPCs across concurrently executing kernels, does not account for contention in the shared L2 cache and memory subsystem.

Although heuristics help guide scheduling decisions, the lack of accurate kernel- or workload-level interference prediction makes it impossible for these schedulers to give accurate performance guarantees. For example, when colocating an instance of GEMV from cuBLAS with different instances of cuBLAS GEMM or CUDNN Conv, on an RTX-3090 using separate SMs per kernel, the slowdown
of GEMV due solely to inter-SM interference can vary from 1.02$\times$ to 2.15$\times$. Existing GPU schedulers lack mechanisms to model or predict the degree of slowdown in each specific colocation scenario. As a result, they cannot distinguish between cases that cause minimal interference and those that lead to severe performance degradation, potentially resulting in SLO violations and suboptimal scheduling decisions. 


\subsection{Existing interference predictors}



Existing  GPU interference predictors operate at either the workload or kernel level. Workload-level approaches such as Horus~\cite{Yeung22Horus}, Abacus~\cite{Cui21Abacus}, and Mudi~\cite{Chen25Mudi} rely on workload-specific features such as model architecture, layer types, or batch size---limiting their generality. Instead, kernel-level predictors model interference between individual kernels and thus generalize better across applications. However, existing kernel-level predictors remain inadequate. Most kernel-level predictors are developed and evaluated in GPU simulators, relying on performance metrics that are unavailable or difficult to obtain on real GPUs~\cite{Tan23Gpupool, Zhao20HSM, Zhao19Themis}.  Even methods designed for real hardware rely on coarse-grained performance counters that fail to capture the complex interactions underlying kernel interference~\cite{Han24Kace}. Our evaluation in section \ref{sec:mosaic_eval} confirms that current predictors exhibit large prediction errors on modern GPUs, substantially under- or overestimating colocation-induced slowdowns.

Interference prediction has also been studied extensively in CPUs~\cite{Subramanian13Mise, SubramanianASM, Eyerman10SMT, Feliu17SMTPower8, Sanghyun26SMTcheck} but existing techniques do not directly translate to GPUs because they similarly depend on simulation or fine-grained microarchitectural metrics that GPUs do not expose~\cite{SubramanianASM, Subramanian13Mise, Eyerman10SMT}. For example, CPI-stack-based methods assume execution cycles can be decomposed into well-defined stall sources, whereas GPUs’ massive thread-level parallelism and latency hiding make such attribution harder, and GPUs expose only limited microarchitectural state for fine-grained analysis~\cite{Sanghyun26SMTcheck, Feliu17SMTPower8}. 

These limitations motivate the need for \textit{an accurate, hardware-based kernel-level interference predictor} that generalizes across applications while capturing the diverse sources of GPU resource contention.

%% file: content/kernelpredictor.tex
\section{\mosaic: GPU kernel interference predictor}\label{kernel_predictor}

We present \mosaic, a kernel-granularity GPU interference predictor. As shown in Figure \ref{fig:predictor-overview}, \mosaic takes as input two GPU kernels and their isolated execution profiles, collected using selected metrics from Nsight Systems and Nsight Compute. It outputs the expected slowdown of each kernel under colocation. \mosaic supports two execution modes: (1) strict SM separation, where concurrent kernels are assigned to disjoint SMs, e.g. using Green Contexts, and (2) full-GPU execution, where kernels may execute concurrently on the same SMs, as with default CUDA streams.

\mosaic estimates slowdown with four main components: \textbf{Thread-block placement} (\S \ref{tb_placement}) models kernel overlap and block execution in the SMs; \textbf{Block packing} (\S \ref{intra_kernel}) captures the slowdown caused when a kernel executes with a different per-SM block density than in isolation; \textbf{Inter-SM interference} (\S \ref{inter_sm})  captures contention for shared L2 and HBM resources even when kernels occupy disjoint SMs; and \textbf{Intra-SM interference} (\S \ref{intra_sm}) captures contention for resources shared by blocks of different kernels executing on the same SM. \mosaic combines analytical models for structured behaviors with lightweight learned models for the remaining effects. It selectively applies these components based on the execution mode, as described in \S \ref{combining}.

The remainder of this section presents each component, starting with empirical observations across four GPU types (H100, A100, RTX-3090, and V100) and a collection of kernels from Table \ref{table:kernels}, Appendix \ref{appendix}. We then describe how we integrate these observations into \mosaic. We will use kernels $A$ and $B$ from Figure \ref{fig:predictor-overview} as examples in this section.

\subsection{Modeling thread block placement}\label{tb_placement}

Thread-block placement determines when the two kernels overlap, whether their blocks can colocate on the same SM, and when overlap begins. We assume kernel A is scheduled before kernel B. Our model is based on three observations:

\begin{figure}[t!]
    \centering
    \includegraphics[width=\columnwidth]{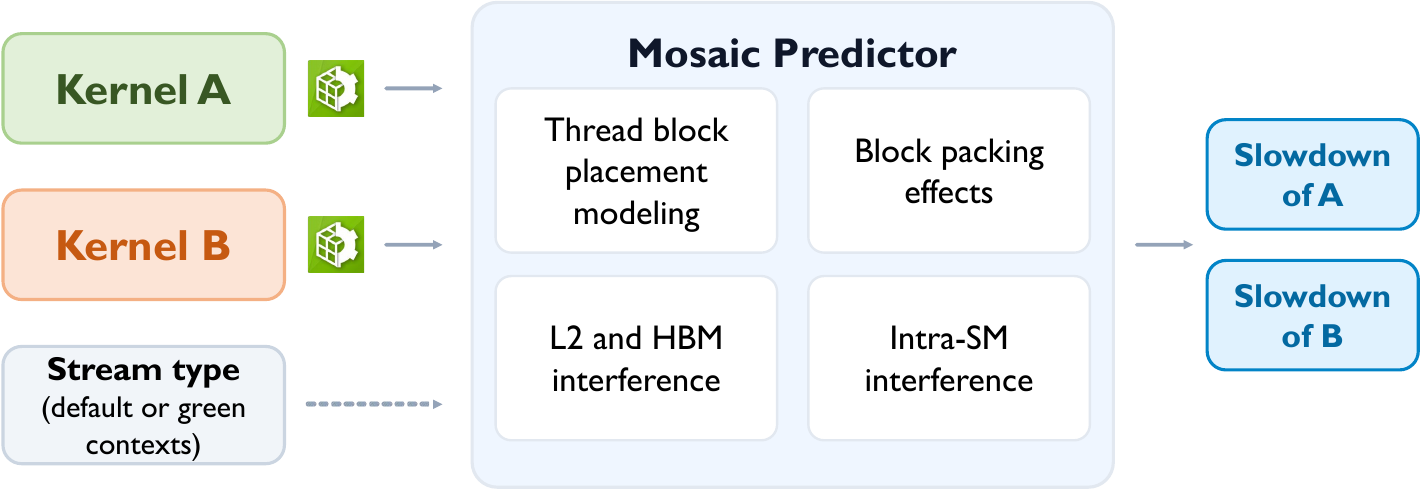}
    \caption{Overview of the \mosaic predictor}
    \label{fig:predictor-overview}
\end{figure}

\textbf{Observation 1:} \textit{When the two streams are of equal priority, the GPU scheduler will first place all blocks from A before any blocks from B.}  If $A$ requires multiple waves, overlap with B can occur only during $A$'s final wave, as earlier waves still contain unscheduled blocks from $A$. Using streams with different priorities might affect this behavior. We leave stream priorities as future work and briefly discuss them in  \S\ref{discussion}. 

\textbf{Observation 2:} \textit{Block placement follows the most-room policy, under SM resource constraints.} Block placement is constrained by available threads, registers, and shared memory. Prior work and our experiments indicate that the scheduler follows a most-room policy~\cite{Gilman20mostroom}, assigning each block to the SM that can accommodate the most blocks from the kernel. L1/shared-memory carveout preferences also matter: incompatible requirements prevent blocks from $A$ and $B$ from sharing the same SM~\cite{L1sharedScheduling}.

\textbf{Observation 3:} \textit{ Accurate placement prediction requires modeling gradual SM resource release.} When all blocks of $A$ have been scheduled, kernel $B$ will start by occupying any idle SMs. If no SMs are available, $B$ must wait until blocks from $A$ complete and resources are released. As resources become available, the scheduler attempts to place blocks from $B$ immediately, potentially packing multiple blocks onto an SM and producing a different placement pattern from the case where all SMs are initially idle.

\textbf{Implications for predictor design:} \mosaic first determines how many blocks from $A$ can fit on each SM and places them using the most-room policy. For $A$’s final wave, we determine the remaining thread-block, register, and shared-memory capacity on each SM. $B$’s blocks are subsequently placed on empty SMs and, when possible, colocated with $A$’s blocks, subject to resource and L1/shared memory carveout constraints. If $B$ cannot be colocated with $A$, its blocks are scheduled as $A$ releases SM resources. To model SM release, \mosaic accounts for variation in $A$'s block execution times. Since block-level execution times are not directly observable for closed-source kernels, we use the \texttt{sm\_\_cycles\_active. \\ avg/min/max} metrics from NCU~\cite{nsightcompute} to approximate the distribution of SM completion times. This allows us to model SM release order and, consequently, $B$’s block scheduling.

\subsection{Modeling block packing}\label{intra_kernel}

During colocation, kernel B may have a different number of concurrently resident blocks per SM than in isolation. Kernel A can occupy some SMs, forcing $B$’s blocks onto the remaining SMs and potentially increasing block density. We therefore model how block density affects $B$’s execution time.

 \textbf{Observation:} \textit{Higher block density increases contention for compute and memory resources within an SM, but the resulting slowdown varies across kernels.} Compute-bound kernels are generally more sensitive, whereas memory-bandwidth-bound kernels can be less affected. Fewer active blocks can reduce memory system contention and improve per-block execution time. Thus, colocation effects depend on both block density and total active blocks.
 
\textbf{Implications for predictor design:} We model $B$’s colocated wave time relative to its isolated wave time using block density and total active blocks.  Let $d_B^{iso}$ and $N^{iso}$ denote $B$’s isolated execution time and number of waves, respectively. We derive $N^{iso}$ from $B$’s launch configuration and the SM resource constraints described in \S\ref{tb_placement}. The time for a wave of $B$ in isolation is $t_w^{iso} = \frac{d_B^{iso}}{N^{iso}}$. Let $k^{iso}$ and $k^c$ denote the number of concurrently resident $B$ blocks per SM in isolation and under colocation, respectively. Similarly, let $X^{iso}$ and $X^c$ denote the total number of active $B$ blocks across all SMs in the two cases. We model the colocated wave time as:
\begin{equation}\label{eq:packing_wave_time}
t_w^c = (\frac{k^c}{k^{iso}})^\beta \cdot t_w^{iso} \cdot (\frac{X^c}{X^{iso}})^\gamma
\end{equation}
The exponent ($\beta$) captures the sensitivity of a kernel to block density changes. ($\beta=0$) implies that wave execution time is unaffected, whereas larger values indicate increasing sensitivity. We model $\beta$ as:
\begin{equation}\label{eq:packing_beta}
\beta = a - b \cdot dram_{b} + c \cdot compute_{max} 
\end{equation}
where $dram_{b}$ is the global memory bandwidth, $compute_{max}$ is the maximum utilization of the GPU compute pipelines, and $a,b,c$ are coefficients that we fit on the collected data per GPU.
The exponent $\gamma$ captures the effect of changing the total number of active blocks. In particular, fewer active blocks can reduce memory-system contention and improve per-block execution time for memory-bandwidth-bound kernels.  We model $\gamma$ as a function of achieved memory bandwidth.

\subsection{Modeling Inter-SM interference}\label{inter_sm}

When kernels execute on separate SMs, they contend for L2 capacity and bandwidth, as well as main memory bandwidth. 
The interference experienced by a kernel depends on both the pressure it places on the memory subsystem and its sensitivity to cache capacity and memory bandwidth. 

\textbf{Observation 1:} \textit{Memory interference is multidimensional.} Aggregate memory throughput is insufficient to characterize inter-SM interference. Modern GPU memory systems exhibit multiple forms of contention, and read and write traffic affect colocated kernels differently.

\begin{figure}[tbp]
    \centering
    \includegraphics[width=\columnwidth]{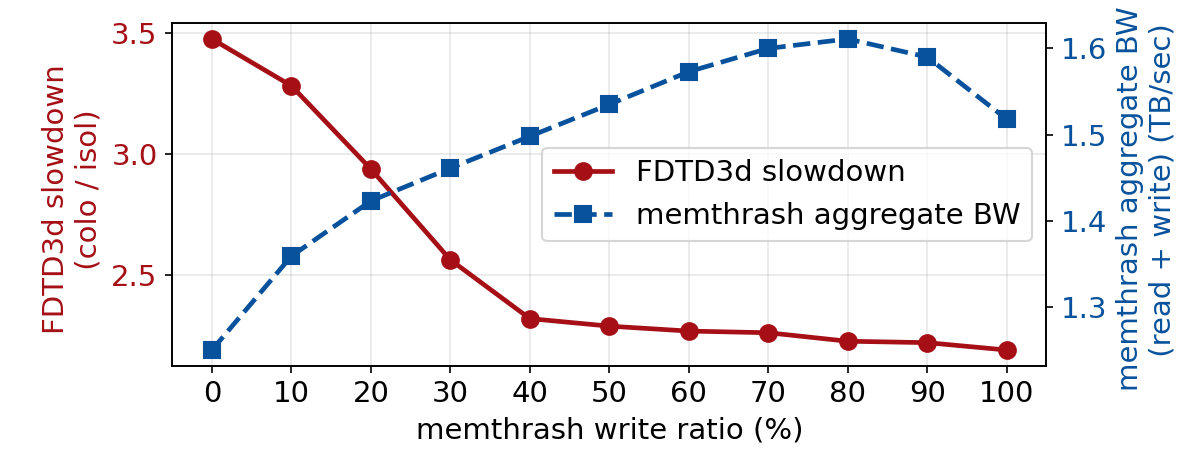}
    \caption{FDTD3d vs a memthrash kernel, colocating in separate SMs on an H100 GPU} 
    \label{fig:inter-read-write}
\end{figure}

For example, Figure \ref{fig:inter-read-write} shows the slowdown of FDTD3d~\cite{cudasamples} when colocated with memThrash, a custom microbenchmark that generates high L2 and global memory pressure while bypassing L1 using inline PTX. We vary memThrash's write ratio from 0\% (read-only) to 100\% (write-only) in 10\% increments, while keeping the working sets of both kernels within L2. We execute FDTD3d and memThrash on 40 and 92 SMs, respectively, on an H100 GPU.
FDTD3d experiences substantially higher slowdown with read vs. write interference. Moreover, the aggregate bandwidth of memThrash increases as the slowdown of FDTD3d decreases, demonstrating that aggregate bandwidth alone does not capture the interference behavior.

\textbf{Observation 2:} \textit{Latency tolerance determines slowdown.} Contention in L2 capacity and bandwidth, and global memory bandwidth, may translate to prolonged memory accesses. The extent to which a kernel tolerates this latency determines its performance degradation. For example, we consider silumul from FlashInfer~\cite{ye2025flashinfer} and FDTD3d~\cite{cudasamples} with similar L2 hit rates, L2 and HBM bandwidth, that are colocated with the same memThrash kernel. silumul experiences a slowdown of 1.3x while FDTD3d a larger slowdown of 2x. silumul has more eligible warps, higher compute-pipeline utilization, and fewer long-scoreboard stalls, indicating greater ability to tolerate increased memory latency.

\textbf{Implications for predictor design:} These observations motivate a feature-based model that captures both memory pressure and latency tolerance. \mosaic distinguishes read and write traffic and includes separate L2 and HBM bandwidth metrics. To capture latency tolerance, we use eligible warps, stall behavior, and relative compute- and LSU-pipeline utilization. \mosaic uses a lightweight XGBoost~\cite{xgboost} model to map the features in Table~\ref{table:inter_sm_metrics} to slowdown. A linear model is insufficient because the interaction between memory pressure and latency tolerance is nonlinear, whereas XGBoost captures these interactions while remaining lightweight enough for practical prediction.

\begin{table*}[tp!] 
\centering
\footnotesize
\begin{tabular}{|c|p{10cm}|}
\hline
\textbf{Metric} & \textbf{Description} \\
\hline
SMs & Number of SMs that the kernel occupies \\
\hline
L1TEX\_L2\_read\_bw, L1TEX\_L2\_write\_bw & Read and write bandwidth between L1TEX and L2 \\
\hline
DRAM\_read\_bw, DRAM\_write\_bw & Read and write bandwidth between L2 and DRAM \\
\hline
Eligible Warps & Average number of warps eligible for execution per cycle \\
\hline
Long Scoreboard stalls & Average number of warps stalled per cycle due to long scoreboard stalls  \\
\hline
Stall ratio & Ratio of non-memory-related stalls  to the long scoreboard stalls \\
\hline
Comp\_to\_global\_lsu\_pipeline & Ratio of compute pipeline usage (Tensor, FMA, XU) to LSU pipeline usage,  scaled by the ratio of shared to global accesses\\
\hline
Sector efficiency & Number of sectors per L2 cache requests \\
\hline

\end{tabular}
\caption{Metrics used to predict inter-sm interference. By L1TEX, we refer to the whole space inside the SM used both for L1 cache and shared memory. We take these metrics (except the number of SMs) from Nsight Compute~\cite{nsightcompute}. }
\label{table:inter_sm_metrics}
\end{table*}

\subsection{Modeling Intra-SM interference}\label{intra_sm}

We now model intra-SM interference, that occurs when blocks from two kernels execute concurrently on the same SM. We identify five major sources of intra-SM interference: (1) execution pipelines, (2) warp scheduling and instruction issue, (3) shared memory, (4) memory instruction (LSU/MIO) units, and (5) L1/L2 cache and memory hierarchy effects. 

\textbf{Observation 1:} \textit{Pipeline and warp-scheduling contention can be modeled analytically using resource overlap and issue intensity.} Modern GPUs expose multiple specialized execution pipelines, making kernels susceptible to contention on the pipelines they use. Contention is strongest when kernels heavily utilize the same pipeline, with kernels having fewer eligible warps experiencing greater slowdown. Memory-bound kernels, however, are largely insensitive to pipeline contention because they spend much of their execution stalled waiting for data.

Colocated kernels also compete for warp-scheduling resources. IPC serves as a proxy for scheduler utilization~\cite{Elvinger26Interference}: a kernel with higher IPC can slow a kernel with lower IPC down. Slowdown generally increases with the IPC difference, although kernels with very low IPC due to global-memory stalls are less affected, as they are already execution-starved.

\textbf{Implications for predictor design: } We model the interference on pipelines and warp scheduler analytically using  $S_{analytical}=max(S_{pipe}, S_{ipc})$. We formulate $S_{pipe}$ as follows:

For each execution pipe $P$ (ALU, FMA, Tensor), let $r^P_A, r^P_B$ be the
$A$'s and $B$'s demand as a fraction of pipe capacity, and
$w_B = n_B/(n_A{+}n_B)$ $B$'s share of resident warps. Then
\begin{equation}
  S_{pipe} =
  \begin{cases}
    1, & \sigma_B > \sigma_{th}, \\[2pt]
    \max\limits_{P}\; \dfrac{r^P_B}{\min\!\big(r^P_B,\ \max(w_B,\, 1-r^P_B)\big)}, & \text{otherwise,}
  \end{cases}
\end{equation}
where $\sigma_B$ is the long scoreboard stalls of $B$ (indicating this kernel is bound by memory accesses), the denominator is the demand $B$ is actually served (which depends on its demand, its warp share and what $A$ leaves unused)

We model $S_{ipc}$ for kernel $B$ as follows ($k$, $k_\sigma$ are constants tuned for each GPU type): \begin{equation}\label{ipc}
    S_{ipc} = 1 + \frac{k}{1 + k_\sigma \cdot \sigma_B} \cdot \max(0, IPC_A - IPC_B)
\end{equation}

\textbf{Observation 2:} \textit{Shared memory and cache subsystems introduce complex interference effects that are difficult to capture accurately using purely analytical models.} Kernels on the same SMs might interfere for shared memory~\cite{Elvinger26Interference}, L1 cache capacity and bandwidth, and the unified memory instruction (MIO/LSU) pipeline, which services both shared and global memory operations. Similarily to the contention for L2 and HBM discussed in section \ref{inter_sm}, these effects are difficult to capture with an analytical model.

\textbf{Implications for predictor design:} We capture these effects using a lightweight XGBoost model that predicts a residual component added to $S_{analytical}$. We use as features L1 and L2 read and write bandwdidth, L1 and L2 hit rate, eligible warps, IPCs, MIO stalls, and shared memory bank conflicts of each kernel. We use the XGBoost model only if both of the contending kernels are non-compute bound, as compute-bound kernels will mostly interfere for the rest of the resources. The output of the XGBoost model, $\Delta_{ML}$ is then applied to $S_{analytical}$, and the overall slowdown is $S = S_{analytical} + \Delta_{ML}$

\subsection{Putting everything together}\label{combining}

Mosaic combines the outputs of the individual models (from  \S\ref{tb_placement}-\S\ref{intra_sm}) to reconstruct the execution timeline of each kernel. \mosaic supports two execution modes: \circled{1} strict SM separation, where concurrent kernels are assigned to disjoint
SMs with mechanisms such as Green Contexts, and \circled{2} full-GPU execution, where kernels may execute concurrently on
the same SMs, as with default CUDA streams. The \mosaic predictor uses different components based on the execution mode. With strict SM separation, kernels overlap from launch and experience only inter-SM interference. Without SM separation, thread-block placement determines when overlap begins and which blocks share an SM, resulting in distinct non-overlap and overlap phases. We describe each mode below, assuming kernels $A$ and $B$ are launched simultaneously on separate streams.

\textbf{Mode \circled{1}:} When kernels $A$ and $B$ run on separate SMs there is no interference for the thread block placement. Since we assume $A$ and $B$ are launched simultaneously, they will also begin execution at the same time. The execution time of $A$ and $B$ when running in isolation in partitioned SMs is $d_A$ and $d_B$ respectively. The duration is pre-profiled, or can be obtained from our packing predictor(\ref{intra_kernel}). $s_{A,B}$ and $s_{B,A}$ denote the inter-SM slowdown factors experienced by $A$ and $B$ due to colocation. Under colocation, the kernels overlap until the first one completes, at time
$t_{ov} = \min(s_{A,B}\, d_A,\, s_{B,A}\, d_B)$. The durations are then
$d'_A = \max\bigl(s_{A,B}\, d_A,\; d_A + (1 - 1/s_{A,B})\, t_{ov}\bigr)$,
and symmetrically for $d'_B$.


\begin{figure}[ht!]
    \centering
    \includegraphics[width=\columnwidth]{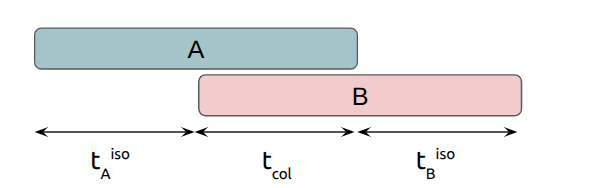}
    \caption{Timeline of two kernels A and B sharing a GPU (with no SM separation). We assume A and B are launched simultaneously, and A is picked first for colocation.}
    \label{fig:mode1-timeline}
\end{figure}

\textbf{Mode \circled{2}:} 
Execution can be divided into three phases: 1) a period during which only kernel $A$ executes, 2) a period during which kernels $A$ and $B$ overlap and interfere, and 3) a period during which only kernel $B$ remains active.  Figure \ref{fig:mode1-timeline} illustrates this timeline. Let $d_A$ and $d_B$ denote the isolated execution times of A and B with all SMs available. A has $n_{{waves}_A}$ waves, with per-wave duration of $w_A$, and B has $n_{{waves}_B}$ waves with per-wave duration of $w_B$. Let $s_{A,B}$ and $s_{B,A}$ denote the slowdown factors experienced by A and B due to colocation. These terms are the max of inter- and intra-SM interference as generated from the inter-SM and intra-SM interference predictors in \ref{inter_sm} and \ref{intra_sm}. For kernel B, the effect of increased block density is captured separately by the intra-kernel packing slowdown $s_B^{pack}$ (\ref{intra_kernel}).  

The thread-block placement predictor (\S\ref{tb_placement}) determines the duration of the first phase, denoted by $t_A^{iso}$. This interval corresponds at least to the execution time of all waves of $A$ except its final wave. Thus, the execution time of kernel $A$ under colocation becomes $dA' = t_A^{iso} + w_A + (s_{A,B}-1) \cdot t_{ov}$, where $t_{ov} = min(w_A, d_B)$.

Suppose that kernel $B$ completes $n_{waves,c}$  waves while overlapping with $A$. During that period, the per-wave time of $B$ will be $s_{B,A} \cdot s_B^{pack} \cdot w_B$. After $A$ has finished, $B$ will run alone, scheduling the remaining $n_{{waves}_B} - n_{waves,c}$ waves in the whole GPU, each with time per wave $w_B$. Thus, overall, we have $dB' = t_A^{iso} +  n_{waves,c} \cdot s_{B,A} \cdot s_B^{pack} \cdot w_B + (n_{{waves}_B} - n_{waves,c}) \cdot w_B $, where we can approximate the waves of $B$ during colocation with $A$ as: $n_{waves,c} = \frac{s_{A,B} \dot w_A}{s_{B,A} \cdot s_B^{pack} \cdot w_B } $.

%% file: content/workloadpredictor.tex
\section{\mosaicsched}

\begin{figure}[tbp]
    \centering
    \includegraphics[width=\columnwidth]{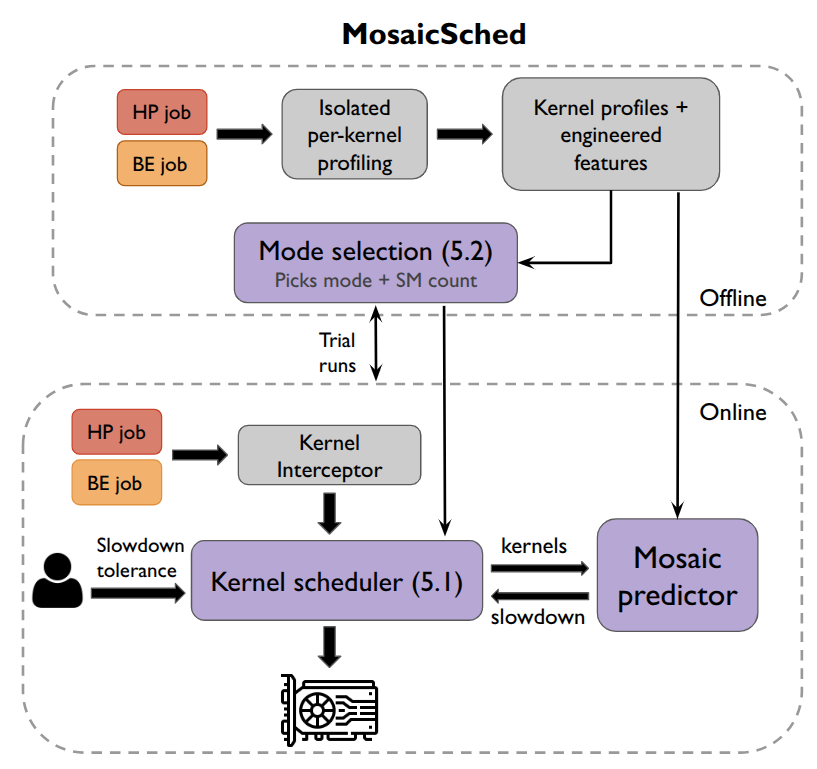}
    \caption{Overview of \mosaicsched.}
    \label{fig:mosaic-integration}
\end{figure}

\S\ref{kernel_predictor} described how Mosaic predicts the execution time of two colocated kernels. We now present \mosaicsched, a GPU scheduler that uses the \mosaic predictor to enable colocating a high-priority (HP) job with a best-effort (BE) job while providing p99 latency guarantees for the high-priority job. \mosaic can be used to build GPU schedulers with different policies and targets, such as enforcing throughput or median latency SLOs. We focus on p99 latency here as tail latency is a key performance objective for GPU colocation and is particularly challenging to control under interference~\cite{Coppock25Lithos, hu2026hummingbird, Strati24Orion, Mingcong22Reef}. As shown in Figure~\ref{fig:mosaic-integration}, \mosaicsched takes as input a set of jobs (where one is designated as HP), a p99 latency SLO expressed a slowdown tolerance \(p\) for the HP job's p99 latency, and profiling information described in \S\ref{kernel_predictor}. The scheduler aims to maximize BE throughput while ensuring that the HP p99 slowdown remains within \(p\). 

We build \mosaicsched on top of Orion~\cite{Strati24Orion}, an open-source, state-of-the-art GPU scheduler that makes runtime, per-kernel scheduling decisions without modifying the underlying models. \mosaic is scheduler-agnostic and can also be integrated with schedulers operating at other granularities, such as thread blocks~\cite{Coppock25Lithos, zhao25Tally}. We retain Orion's interception mechanism but replace its scheduling policy with \mosaic-based admission control, described in 
\S\ref{sched:policy}. The scheduler maintains the execution state of the HP and BE workloads, including completed kernels, in-flight kernels, and their predicted completion times. We use CUDA events~\cite{cudaevents} to track kernel completion.  Furthermore, \mosaicsched selects between having the two jobs sharing the full GPU or enforcing strict SM separation. For the latter, we maintain a pool of streams associated with SM subsets created using Green Contexts; otherwise, the jobs execute on dedicated default CUDA streams\footnote{Despite one of the jobs being high-priority for the client, \mosaicsched uses CUDA streams of equal priorities to submit the kernels 
}. We select between these execution modes using offline profiling, as described in \S\ref{sched:selection}.

\subsection{GPU scheduling with \mosaic-enabled predictions}\label{sched:policy} 

\mosaicsched dispatches HP kernels immediately in the associated stream. When a BE kernel $b$ arrives, the scheduler identifies the HP kernels expected to overlap with $b$ and invokes the \mosaic predictor to predict their execution times under the selected execution mode. Kernel $b$ is admitted only if the predicted completion time of the high-priority request remains within the user-specified slowdown target. We admit $b$ if:
\begin{equation}
    T_{completed} +T_{inflight} + T_{overlap(b)}+T_{remaining}  \leq p' \cdot T_{iso},
\end{equation}
where $T_{completed}$ denotes the execution time accumulated by completed high-priority kernels, $T_{inflight}$ is the expected duration of in-flight kernels based on the
scheduling decisions so far, $T_{overlap(b)}$ is the predicted execution time of the high-priority kernels expected to overlap with $b$, and $T_{remaining}$ corresponds to the execution time of the yet-to-be-launched high-priority kernels of this request  assuming no colocation. $T_{iso}$ is the isolated execution time of the entire high-priority request, and $p'$ is the applicable slowdown budget. In full-GPU mode, \(p'=p\), the user-specified p99 slowdown tolerance. Under strict SM separation, the scheduler accounts separately for the slowdown incurred by restricting the HP workload to its assigned SMs.

\subsection{Selecting the GPU execution mode}\label{sched:selection} 

\mosaicsched chooses between full-GPU sharing and strict SM separation to maximize BE throughput while remaining below the HP p99 slowdown tolerance. We perform this selection offline using short profiling runs for candidate configurations. We test full-GPU execution, where we run the scheduler using the user-specified slowdown tolerance $p$ and measure BE throughput. We also test SM separation, where we first use Mosaic's packing model (\S\ref{intra_kernel}) to estimate the HP execution time for candidate SM allocations and discard allocations whose isolated slowdown already exceeds $p$. For each remaining allocation, we run the scheduler with adjusted slowdown tolerance with the corresponding Green Context configuration and measure BE throughput. \mosaicsched selects the mode and SM allocation with the highest BE throughput. We restrict candidate SM allocations to multiples of eight to reduce the search space. Although the scheduler supports different allocations across consecutive HP kernels, in this paper, we use a fixed allocation to avoid the overhead of frequent Green Context transitions. 

\subsection{Minimizing prediction overheads}\label{sched:overheads} 
To reduce the overhead of computing predictions, we precompute engineered features, compile the XGBoost trees into optimized C++, and preallocate fixed-size feature vectors. These optimizations reduce prediction latency to approximately \(1\,\mu s\) for inter-SM prediction and \(7-8\,\mu s\) for full-GPU prediction. Since kernels in ML workloads are repeated, we further reduce the prediction time by \textit{caching} the predictions for unique pairs of kernels.

%% file: content/eval.tex
\section{Evaluation}\label{sec:mosaic_eval}

We evaluate \mosaic and \mosaicsched to answer:

\begin{itemize}
    \item Can \mosaicsched satisfy user-defined high-priority SLOs while maximizing best-effort throughput, compared with baselines?
    \item What are the cost benefits of GPU colocation?
    \item How accurate is the \mosaic predictor?
    \item How important is each component of  \mosaic?
\end{itemize}

\subsection{Methodology}\label{sec:mosaic_eval_methodology}

\textbf{Experimental Setup:} We evaluate the \mosaic predictor in four GPU types: H100, and RTX-3090 in on-premise machines, and A100 and V100 GPUs in Google Cloud (using \texttt{a2-highgpu-1g} and \texttt{n1-standard-8} VMs respectively). We evaluate \mosaicsched in our H100 GPU.

\textbf{Predictor baselines:} We compare prediction accuracy with two state-of-the-art predictors: KACE and GPUPool~\cite{Han24Kace, Tan23Gpupool}. KACE uses Nsight Compute metrics with linear regression. GPUPool was implemented on GPGPU-Sim and uses simulator-specific metrics. For real-GPU evaluation, we collect the closest corresponding Nsight Compute metrics and train its Gradient Boosting regressor using the parameters reported in the original paper.

\textbf{GPU scheduler baselines:} We compare with default CUDA streams, a Green-Context baseline, and Orion~\cite{Strati24Orion}. For the Green-Context baseline, we create streams with disjoint GPU SM sets and assign the HP job the minimum number of SMs required so that its slowdown due to SM constraining is less than 1.1x. We use this baseline instead of LithOS~\cite{Coppock25Lithos} because LithOS is not publicly available. This static, HP-favorable allocation follows LithOS’s per-kernel right-sizing principle while avoiding the overheads of Green Context switching between kernels. Because SM constraining contributes negligible slowdown to the HP job, any additional slowdown reflects the inter-SM interference.

\textbf{Workloads:} We use the kernels in Appendix \ref{appendix}, Table \ref{table:kernels} to train and test \mosaic and the baselines, KACE and GPUPool. For workload GPU colocation, we use ResNet50 and Mobilenet-v3~\cite{torchvision}, DLRM~\cite{dlrm}, BERT~\cite{bert} and LLAMA-3-8B~\cite{llama3-8b-instruct, Kwon2023VLLM}, all performing inference with fixed batch sizes. These models span different architectures, generations, and kernel types (convolutions, attention, etc.), demonstrating that \mosaic and \mosaicsched are architecture-agnostic.

\subsection{\mosaicsched evaluation}

Figure \ref{fig:hp-be-colocation} shows the p99 latency slowdown of the HP job and the aggregate throughput of the HP and BE jobs under GPU colocation. Each bar corresponds to a HP job labeled on the x-axis, with results averaged across experiments pairing it with different BE jobs. The error bars show the min and max slowdowns across the different BE runs. The slowdown is computed as the ratio of the colocated p99 latency to the p99 latency of the HP job in isolation. We additionally plot the sum of isolated HP and BE throughputs, as the best achievable throughput in the absence of colocation interference.

\subsubsection{SLO satisfaction and comparison with baselines}
\begin{figure*}[ht!]
    \centering
    \begin{subfigure}{\textwidth}
        \centering
        \includegraphics[width=\textwidth]{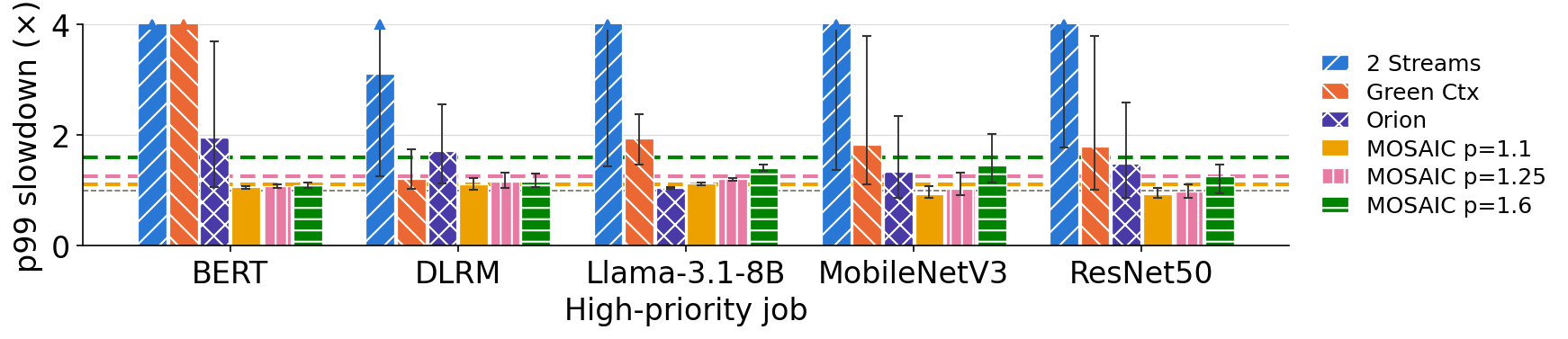}
        \caption{HP tail latency (p99) slowdown under colocation. For \mosaic, $p$ stands
        for the user-defined slowdown tolerance. Bars show the mean over BE jobs, and the error lines show the min and max p99 latencies across the BE pairs. The Isolated p99 latencies of the HP jobs are: 108ms for BERT, 30ms for DLRM, 1.5sec for LLAMA, 14ms for MobileNet, 12ms for ResNet50.}
        \label{fig:hp-be-colocation-latency}
    \end{subfigure}
  
    \begin{subfigure}{\textwidth}
        \centering
        \includegraphics[width=\textwidth]{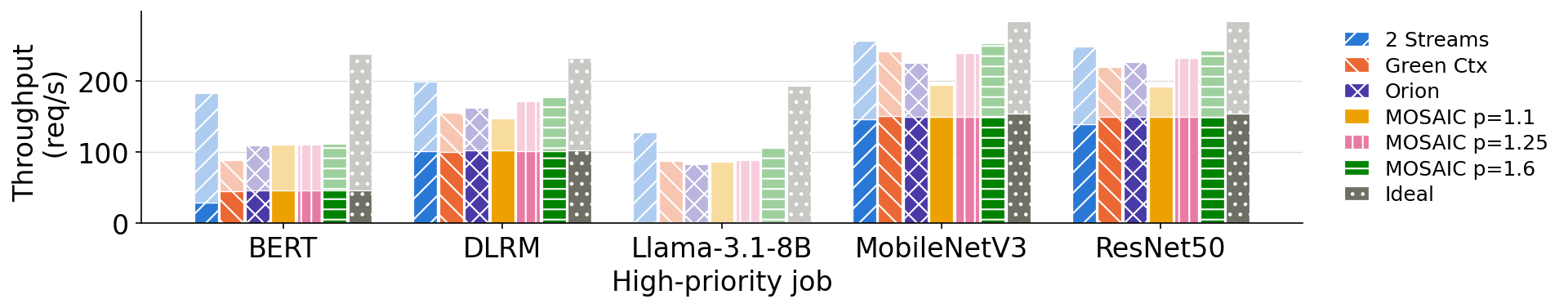}
        \caption{System throughput under colocation, stacked as HP job (bold)
        $+$ BE job (faint).}
        \label{fig:hp-be-colocation-throughput}
    \end{subfigure}

    \caption{HP/BE job colocation: p99 latency slowdown (top) and
    throughput~(bottom). \emph{Ideal} is the isolated single-job run. 
    }
    \label{fig:hp-be-colocation}
\end{figure*}

Figure \ref{fig:hp-be-colocation-latency} compares the p99 latency slowdown of each HP job across \mosaicsched (for different user-defined penalties) and the baselines. Uncontrolled colocation using two default CUDA streams causes significant HP slowdown, with p99 latency slowdown being larger than 10$\times$ for several workloads. Orion reduces this degradation by selectively admitting BE kernels, but its coarse-grained interference model can still make suboptimal admission decisions, resulting in HP slowdowns of up to 3.7×. Strict SM separation alone also does not eliminate interference. The Green Context baseline incurs 1.1–7$\times$ slowdown due to the contention for shared L2 and HBM resources, despite 
the SM partitioning favoring the HP job.

MosaicSched consistently keeps HP p99 slowdown at or near the specified tolerance across workloads and SLO targets, by jointly selecting the execution mode and SM allocation and performing kernel-level admission control using its interference predictor. We quantify SLO violation as the fraction of requests whose p99 latency exceeds the user-specified slowdown tolerance. Across most workloads, MosaicSched satisfies the p99 SLO, with at most 1.0\% of requests exceeding the SLO latency target. The only exception is MobileNet at p=1.6 where  1.2\%  of request exceed the SLO, due to prediction errors for a small number of kernel pairs, which can be amplified by subsequent queuing. In contrast, violations are more frequent for all baselines and become more pronounced as the SLO becomes stricter. For (p=1.1), the 2-Streams, Orion, and Green Context baselines violate the SLO for 26–99\%, 0.9–14\%, and 2.0–54\% of requests, respectively. At (p=1.25), these rates are 23–99\%, 0.1–12.1\%, and 1.3–49\%, and at (p=1.6), 17–99\%, 0.1–9.0\%, and 0.4–39\%, respectively.

\subsubsection{Throughput and cost-benefit analysis}

Figure \ref{fig:hp-be-colocation-throughput} shows the throughput of the HP job (bold bar) and BE job (faint bar). Uncontrolled colocation using default CUDA streams maximizes BE throughput at the expense of HP throughput. Other baselines limit BE throughput through SM partitioning (Green Contexts) or BE kernel admission control (Orion). For \mosaicsched, BE throughput increases monotonically with the HP slowdown tolerance (p), as greater SLO slack provides more flexibility to schedule BE kernels. At moderate penalties (p=1.25 and p=1.6), \mosaicsched achieves higher BE throughput than Orion while maintaining lower HP p99 latency, demonstrating the benefit of more accurate and controlled BE kernel admission.

Table \ref{tab:cost-savings} shows the cost benefits of colocating best-effort with high-priority jobs from Figure \ref{fig:hp-be-colocation}, using \mosaicsched under different penalties. For a pair of BE-HP jobs, the cost savings are coming from packing the two jobs in one GPU, rather than dedicating each job to a GPU. Similar to Orion~\cite{Strati24Orion}, we calculate the cost savings as $2 \cdot \frac{Thr_{BE}^{colocated}}{Thr_{BE}^{isolated}} $
where $Thr_{BE}^{isolated}$ is the BE throughput when running alone, and $Thr_{BE}^{colocated}$ is the BE throughput when colocated with an HP job. A value below 1 indicates that colocation provides no cost benefit, as the BE job achieves less than half its isolated throughput. The maximum value of 2 corresponds to negligible BE slowdown and thus maximal cost benefit.

Cost savings vary substantially across BE workloads and HP SLO slack. Under a tight slowdown tolerance (p = 1.1), \mosaicsched has limited flexibility to schedule BE kernels without violating the HP SLO, often severely throttling BE throughput and providing little or no cost benefit. As the tolerance increases, \mosaicsched can allocate more resources to the BE job, making colocation increasingly beneficial. At p = 1.6, all workloads achieve mean cost savings above 1. The variation across BE workloads further highlights that the benefit of colocation depends on the interaction between the colocated jobs. Running \mosaicsched for a few rounds on a pair of HP and BE jobs helps determining whether colocation is cost-beneficial or not.

\subsubsection{Full-GPU sharing vs. SM separation} \mosaicsched decides whether to let the two jobs share the GPU SMs or constrain them with Green Contexts. This choice depends on both the workload pair and the HP slowdown tolerance. For example, in Figure \ref{fig:hp-be-colocation}, \mosaicsched selects full-GPU sharing for most vision workloads, while it favors SM separation with 64–96 SMs allocated to the HP job for DLRM. Under low slowdown tolerance, full-GPU sharing is preferable because \mosaicsched can selectively admit BE kernels while preserving the HP SLO. Indeed, for \(p=1.1\), \mosaicsched selects full-GPU sharing for all workloads in Figure \ref{fig:hp-be-colocation}.

SM separation introduces a tradeoff between reducing interference and providing sufficient resources to the BE job. Allocating more SMs to the HP job reduces interference and allows Mosaic to admit more BE kernels, but leaves fewer resources for those kernels. Conversely, allocating fewer SMs to the HP job increases interference but provides more resources to each admitted BE kernel. Thus, admitting more BE kernels does not necessarily yield higher BE throughput, and the full-GPU mode can outperform SM-seperation mode despite admitting fewer kernels.

The impact of SM partitioning and kernel admission also depends on kernel characteristics. Compute-bound BE kernels typically cause less memory-system interference and are admitted more frequently, but can suffer substantially from SM reduction. Memory-bound kernels are generally more resilient to SM reduction but create greater L2 and memory contention, reducing their admission rate. Consequently, no single execution mode consistently maximizes BE throughput while meeting the HP latency target. 
Mosaic therefore performs a few profiling runs (\S\ref{sched:selection}) to select the best mode for each job pair and latency target.

\begin{table}[ht!]
    \centering
    \small
    \begin{tabular}{lrrr}
        \toprule
        BE job & $p{=}1.1$ & $p{=}1.25$ & $p{=}1.6$ \\
        \midrule
        \texttt{ResNet50} & 0.00/0.27/0.92 & 0.67/1.06/1.33 & 0.69/1.13/1.50 \\
        \texttt{MobileNet} & 0.64/0.98/1.18 & 0.64/1.14/1.44 & 0.64/1.24/1.59 \\
        \texttt{BERT} & 0.89/1.05/1.18 & 0.94/1.14/1.25 & 1.01/1.31/1.56 \\
        \texttt{DLRM} & 0.00/0.12/0.70 & 0.72/1.14/1.67 & 0.73/1.21/1.68 \\
        \texttt{Llama} & 0.04/0.04/0.07 & 0.04/0.20/0.85 & 0.90/1.12/1.29 \\

        \bottomrule
    \end{tabular}
     \caption{Best-effort cost savings with \mosaic under different penalty constraints. Each entry reports min/mean/max cost savings across workloads. Cost savings are relative to using two separate GPUs; values below 1 indicate no cost savings. The maximum possible value is 2.}\label{tab:cost-savings}
\end{table}

\subsection{\mosaic Predictor accuracy}\label{sec:mosaic_eval_error} 

We evaluate \mosaic’s prediction accuracy against KACE and GPUPool, with and without SM partitioning, by measuring kernel-pair slowdowns under colocation. To assess generalization in unseen kernels, we train each model on a subset of Table \ref{table:kernels} kernels with varying input sizes, launch parameters, and SM partitioning, and evaluate on unseen kernels. 
Note that we retrain \mosaic and the baselines in each different GPU type. Reducing the amount of retraining required across GPU generations is left for future work.

\subsubsection{With SM partitioning}\label{sec:mosaic_eval_error_inter_sm}

This experiment corresponds to Mode \circled{1} of the \mosaic predictor, where colocated kernels execute on separate SMs using Green Contexts. Table \ref{table:error_inter_sm} reports the average p50 and p95 absolute percentage errors of \mosaic and the baselines on unseen kernels.

\mosaic reduces p50 error by 3.2–6.3× vs. KACE and 2.2–4.0× vs. GPUPool, and p95 error by 2.1–3.5× and 1.45–3.0×, respectively. 
KACE relies on linear regression and cannot capture inter-SM interference in the cache and memory subsystems. GPUPool uses aggregate metrics  such as global memory bandwidth, pipeline utilization, IPC, and memory requests but does not distinguish read/write bandwidth or account for latency hiding. Hyperparameter tuning only slightly improves its H100 p95 error, from 45\% to 44\%. This highlights the importance of accurate features and explains Mosaic’s lower error on unseen kernels.

\begin{table}[t]
    \centering
    \begin{tabular}{lcccc}
        \toprule
        \textbf{Predictor} & \textbf{H100} & \textbf{RTX} & \textbf{A100} & \textbf{V100} \\
        \midrule
        KACE   & 19 / 51 & 23 / 52 & 19 / 54 & 38 / 101 \\
        GPUPool & 11 / 45 & 24 / 72 & 24 / 58 & 15 / 72 \\
        Mosaic & \textbf{5 / 23} & \textbf{6 / 24} & \textbf{6 / 26} & \textbf{6 / 27} \\
        \bottomrule
    \end{tabular}
    \caption{Average p50/average p95 absolute percentage errors (\%) when kernels run on seperate SMs.}
    \label{table:error_inter_sm}
 \end{table}

\subsubsection{Without SM partitioning}\label{sec:mosaic_eval_error_full}

Table \ref{table:error_full} reports p50 and p95 absolute percentage errors for unseen kernels under default CUDA streams, where kernels may share GPU SMs. Prediction errors are higher than in the inter-SM case (\ref{sec:mosaic_eval_error_inter_sm}) because shared SMs introduce contention for thread-block scheduling and intra-SM resources in addition to inter-SM resources. 
Limited SM resources, differing L1/shared-memory carveouts, and warp-scheduling contention can cause substantial slowdowns.
Since neither KACE nor GPUPool explicitly models these effects, both methods frequently and significantly underestimate kernel slowdowns. 
In contrast, \mosaic achieves substantially lower errors by explicitly modeling thread-block scheduling, block packing, and intra- and inter-SM interference, reducing p50 and p95 errors by 7–82× across GPUs. 

\begin{table}[tp!]
    \centering
        \small

    \begin{tabular}{lcccc}
        \toprule
        \textbf{Predictor} & \textbf{H100} & \textbf{RTX} & \textbf{A100} & \textbf{V100} \\
        \midrule
        KACE    & 286 / 1181 & 424 / 3350 & 212 / 980 & 112 / 1288 \\
        GPUPool & 169 / 1482 & 203 / 1431 & 129 / 1393 & 78 / 781.6 \\
        \mosaic & \textbf{8 / 48.2} & \textbf{6 / 41} & \textbf{8 / 38} & \textbf{11 / 49.5} \\
        \bottomrule
    \end{tabular}

    \caption{Average p50/average p95 absolute percentage errors (\%) when kernels are allowed to share  GPU SMs 
    }
    \label{table:error_full}
\end{table}

\subsection{Predictor ablation study}

\textbf{Full Predictor:} Table \ref{table:ablation-full} shows the ablation study for the different components of the \mosaic predictor. Accurate thread-block placement (\S\ref{tb_placement}) alone achieves substantially lower prediction error than KACE and GPUPool by capturing kernel overlap in space and time.  Adding the block-packing model (\S4.2) further reduces p50 error from 15\% to 8.7\%. Finally, incorporating the intra- and inter-SM predictors (\S4.3–4.4) reduces p50 error to 8.0\% and p95 error  to 48.2\%, highlighting their importance for reducing tail prediction errors.

\begin{table}[t]
  \centering
  \begin{tabular}{lcc}
    \toprule
    \textbf{Stage} & \textbf{p50 $|$err$|$ (\%)} & \textbf{p95 $|$err$|$ (\%)} \\ \midrule
    TB placement  & 15 & 68 \\ 
    +Block packing      & 8.7 & 64 \\ 
    +intra + inter (full) & 8.0 & 48.2 \\ 
    \bottomrule
  \end{tabular}
    \caption{Ablation study for the \mosaic predictor in the full GPU mode on the kernels from Table \ref{table:error_full} for the H100 GPU.}
  \label{table:ablation-full}
\end{table}

\textbf{Intra-SM Predictor:} Table 7 evaluates the intra-SM predictor components on setups where kernels execute concurrently on the same SMs. We ablate the analytical pipeline/warp-scheduling model (No Pipe/IPC) and the XGBoost memory model (No Mem), grouping kernels by their resource bottleneck. The analytical component is most important for compute-heavy kernels, increasing p95 error from 46.8\% to 80.0\% when removed, while the XGBoost component is most important for global- and shared-memory-heavy kernels. Thus, the two components capture complementary sources of intra-SM interference. 

\begin{table}[t]
  \centering
    \begin{tabular}{cccc}
    \toprule
    \textbf{Type} & \textbf{Full intra} & \textbf{No pipe/IPC}  & \textbf{No Mem} \\ \midrule
    compute heavy       & 46.8 & 80.0 & 46.8 \\ 
    global-mem heavy     & 61.6 & 88.1 & 84.7 \\ 
    shared mem heavy & 35.2 & 42.7 & 75.7 \\ 
    \bottomrule
    \end{tabular}
     \caption{Ablation study for the \mosaic intra-sm predictor on different kernel types. The table shows the p95 error (\%). }
       \label{table:ablation-intra}
\end{table}

\textbf{Inter-SM Predictor:} For the inter-SM part of the \mosaic predictor, feature selection is more important than XGBoost hyperparameters:
restricting the model to memory and L2 bandwidth metrics only increases H100 p95 error from 23\% to 44\%.

%% file: content/discussion.tex
\section{Discussion}\label{discussion}

We discuss current limitations of the \mosaic predictor (\S\ref{discussion:limitations})     and applicability to GPUs from other vendors (\S\ref{discussion:AMD}).

\subsection{Limitations of the \mosaic predictor}\label{discussion:limitations}


\textbf{Power throttling and frequency deviations:} \mosaic does not explicitly model the effects of power throttling, which can occur under spatial GPU co-location and reduce GPU frequency, thereby increasing kernel execution time. 
Our experiments show that throttling is relatively infrequent but varies across GPU generations and systems (e.g., on-premise H100 clusters vs cloud-based V100 instances),
While some throttling effects may be implicitly captured by our XGBoost learned models, further experimentation is necessary to model these effects.

\textbf{No support for CUDA stream priorities:} The thread block placement predictor (\S\ref{tb_placement}) assumes equal CUDA stream priorities. Stream priorities can influence scheduling by favoring thread blocks from higher-priority streams when they are available, thereby altering block interleaving and execution time. Supporting priorities is future work.

\textbf{GPU-specific model calibration:} The inter- and intra-SM XGBoost models and analytical parameters (\S\ref{intra_sm}) are calibrated per GPU, so supporting new GPU generations requires retraining. Reducing this overhead is future work.

\subsection{Applicability to AMD GPUs}\label{discussion:AMD}
Although we focus on NVIDIA GPUs, we have also experimented with AMD GPUs. Several \mosaic components transfer directly. For example, L2-cache and memory interference can be modeled accurately using XGBoost and existing features. However, AMD’s greedy thread-block scheduler can mix blocks from different kernels, unlike the behavior observed in \S\ref{tb_placement}. Thus, extending \mosaic to AMD GPUs is feasible but requires adapting its scheduling model to vendor-specific behavior.

%% file: content/conclusion.tex
\section{Conclusion}

We presented \mosaic, an accurate and lightweight kernel-level predictor for GPU interference. \mosaic combines analytical models for thread block scheduling, warp scheduling, and compute pipeline contention with xgboost models for the memory subsystem, achieving up to an order of magnitude
lower p50 and p95 error than prior predictors while keeping per-prediction latency in the microsecond range. Using \mosaic, we built \mosaicsched, a GPU scheduler for workload colocation. We  demonstrated that it enables explicit, user-defined performance guarantees for high-priority jobs while admitting best-effort work whenever headroom allows, improving GPU utilization with reliable guarantees.

%% file: content/appendix.tex
\appendix

\section{Appendix}\label{appendix}

\begin{table*}[htp!] 
\centering
\footnotesize
\begin{tabular}{|c|c|c|c|c|c|}
\hline
\textbf{Kernel} & \textbf{Source} & \textbf{Main Pipelines} & \textbf{Mem BW Usage} & \textbf{Shared memory Usage} & \textbf{Main Stall Reason} \\
\hline
Copy & \cite{Elvinger26Interference} & LSU/ALU & high & none & Long scoreboards \\
\hline
Gemm & cuBLAS & FMA/TCORES & low  & heavy & Not selected/Short Scoreboards \\
\hline
Gemv & cuBLAS & LSU/FMA/TCORES & high & moderate & Long Scoreboards\\
\hline
Softmax & cuDNN & ALU/ADU/FMA & high & None & Long Scoreboards \\
\hline
FDTD & cudaSamples & LSU/FMA & low & high & Long Scoreboards/Barrier \\
\hline
Histogram & cudaSamples & LSU/ALU & low & high (atomics) & Long Scoreboards/MIO \\
\hline
Mul & \cite{Elvinger26Interference} & FMA & low & None & Not selected/math \\
\hline
Pchase & custom & LSU/ALU & low & None & Long Scoreboards \\
\hline
Scan & cudaSamples & ADU/LSU/ALU & high & high & Barrier/Branch/Long Scoreboards \\
\hline
Transpose & cudaSamples & LSU/ADU/ALU & low & high & Barrier/MIO/Long Scoreboards \\
\hline
Attention & FlashInfer &  FMA/LSU & low & high & Short/Barrier/Long Scoreboards\\
\hline
SiluMul & FlashInfer & ALU/XU & high & high & Barrier/Short/Long scoreboards \\
\hline
Convolution & cuBLAS & ADU/FMA/TCores & low & high & Short scoreboards/Wait \\
\hline
BatchNorm & cuBLAS & ALU/LSU & medium & none & Long scoreboards\\
\hline
SpMV & custom & LSU/FMA/ALU & low & None & Long Scoreboards \\
\hline
RmsNorm & FlashInfer & ALU/FMA & medium & high & Long Scoreboards \\
\hline
\end{tabular}
\caption{List of the different kernels used for training and testing the \mosaic predictor. We use kernels from ~\cite{Elvinger26Interference}, cuBLAS~\cite{cublas}, cuDNN~\cite{cudnn} and FlashInfer~\cite{ye2025flashinfer}, and custom microbenchmarks.}
\label{table:kernels}
\end{table*}

As discussed in \S\ref{kernel_predictor}, we analyzed interference behavior in four GPU types (H100, A100, RTX-3090, and V100), using a variety of kernels, with different resource requirements, size and execution time.  Creating a complete and diverse dataset is a crucial step towards building an accurate and robust interference predictor. Table \ref{table:kernels} shows the kernels we use for our dataset. We cover 4 main kernel categories: 1) memory-bandwidth bound, 2) compute-bound, 3) global-memory-latency intensive, and 4) shared-memory intensive. We take kernels from ~\cite{Elvinger26Interference}, cuda Samples~\cite{cudasamples}, cuDNN~\cite{cudnn}, cuBLAS~\cite{cublas} and FlashInfer~\cite{ye2025flashinfer}, and also include some custom microbenchmarks. Varying the input size and parameters, we change the resource requirements of the kernels, making them less or more sensitive to different types of interference.